\documentclass[%
aip,
jmp,%
amsmath,amssymb,
reprint,%
]{revtex4-1}

\usepackage{graphicx}% Include figure files
\usepackage{dcolumn}% Align table columns on decimal point
\usepackage{bm}% bold math
\usepackage{cleveref}
\usepackage{url}
\usepackage[usenames,dvipsnames,svgnames,table]{xcolor}
\usepackage[T1]{fontenc}
\usepackage[english]{babel}
\begin{document}
	
	%\preprint{AIP/123-QED}
	
	\title[Filamentous packings]{Perspective: Mechanics of randomly packed filaments -- the `bird nest' as meta-material}
	
	\author{N. Weiner}
	\thanks{These authors contributed equally.}
	\affiliation{Department of Polymer Science, University of Akron.}
	\author{Y. Bhosale}
	\thanks{These authors contributed equally.}
	\affiliation{Department of Mechanical Science and Engineering, University of Illinois at Urbana-Champaign}
	\author{M. Gazzola}
	\email{mgazzola@illinois.edu}
	\affiliation{Department of Mechanical Science and Engineering, University of Illinois at Urbana-Champaign}
	\affiliation{National Center for Supercomputing Applications, University of Illinois at Urbana-Champaign}
	\author{H. King}
	\email{hking@uakron.edu}	
	\affiliation{Department of Polymer Science, University of Akron.}
	\affiliation{Department of Biology, University of Akron.}
	\affiliation{Biomimicry Research and Innovation Center, University of Akron.}
	\date{\today}% It is always \today, today, but any date may be explicitly specified
	\begin{abstract}
%\textcolor{red}{Systems of randomly packed, macroscopic elements, from jammed grains and tangled filaments to crumpled sheets, display mechanical responses which are fundamentally different from, yet strongly dependent upon, those of their constituent parts.  These packings can be thought to represent a broad class of metamaterials, with properties that robustly emerge from disorder, rather than being prescribed by regular structure.  We will present a high-level review of basic research connecting random structure to mechanics, illustrate examples of the fundamental idea manifested in biological systems, and highlight directions of impactful application.}
Systems of randomly packed, macroscopic elements, from jammed spherical grains to tangled long filaments, represent a broad class of disordered meta-materials with a wide range of applications and manifestations in nature. A `bird nest' presents itself at an interface between hard round grains described by granular physics to long soft filaments, the center of textile material science.  All of these randomly packed systems exhibit forms of self assembly, evident through their robust packing statistics, and a common, unusual elastoplastic response to oedometric compression.  In reviewing packing statistics, mechanical response characterization and consideration of boundary effects, we present a perspective that attempts to establish a link between the bulk and local behaviour of a pile of sand and a wad of cotton, demonstrating the nest's relationship with each. Finally, potential directions for impactful applications are outlined. 

	\end{abstract}
	
	\pacs{Valid PACS appear here}% PACS, the Physics and Astronomy
	% Classification Scheme.
	%Use showkeys class option if keyword
	%display desired
	\maketitle

\section{\label{sec:level1}Introduction}
\subsection{Observation of a nest}
A cardinal uses its own body as template in building its cup-nest.  % (Fig \ref{fig:nest}). 
Found, filamentous materials are added and randomly packed against the bird-defined boundaries. The resulting structure is not strong, and one instinctively handles it delicately. Yet we know this structure has been given the profound responsibility of protecting the bird’s offspring. If defined as a random packing of elastic filaments, the bird nest is an unusual material: it is cohesive without attractive interactions; it is plastic although its elements are elastic; it is soft while its filaments are not. It embodies an instinctive understanding of granular mechanics, and yet it combines long flexible elements, impermanent frictional contacts and boundary effects in a way that notably sets it apart from classical grain systems, as well as semiflexible polymer networks and other non-woven materials which derive mechanical response from permanent crosslinks.
How did this solution arise, what can we learn from it, and can it be usefully applied in an engineering context?

\begin{figure}[h!]
	\includegraphics[width=0.4\textwidth]{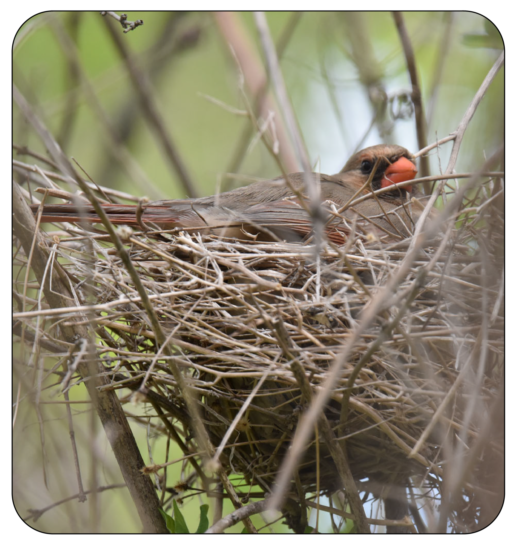}
	\caption{A cardinal nest \cite{cardinal} easily supports the weight of the bird, as well as other perturbations: a cohesive granular structure made up of flimsy elastic filaments.}
	\label{fig:nest}
\end{figure}
		
\subsection{The nest as a construction template for novel materials}

The history of changing environmental pressures has driven birds to their diversity of present day nesting solutions. A cooling climate led most to incubate with their own body heat; predation led many to do so in hidden locations up in the trees; and competition for convenient sites led to fabrication with lightweight (transportable by flight) materials\cite{nest_evolution}. Among the tree-nesting birds, some have spent millions of years working with fibers to build structures for protection and microclimatic regulation\cite{design-funct,collias1997origin}. Several needs drive bird nest design across the diversity of nests, but structural integrity under mechanical loads and disturbances, over its lifetime, is clearly a dominant factor. Large platform nests of eagles and hawks appear to derive stability from gravitational load of heavy sticks. Hummingbirds are known to use spiderweb as sticky lashing, and weavers learn to tie formal knots. Many birds, though, seem to rely on a fundamentally different strategy, rooted in the emergence of desirable properties of the random packing. The additional stick, selected based on some mechanical criteria, does not serve to balance a torque or support an anticipated load, but rather contributes to the target material behavior of the aggregate. In doing so, they effectively apply nuances of scientific principles that humans are only beginning to understand, and without need for abstraction, mechanically synthesize multifunctional metamaterials to suit their needs. This is in stark contrasts with a typical human construction approach: with the same starting materials, the engineer would naturally see a collection of objects, of predetermined material properties, and seek to bond them to each other in a prescribed `structure made of wood'.  Our \emph{intelligent}, prescriptive design process has proven successful, but could only stand to benefit by incorporating strategies of \emph{emergent} design.

\begin{figure}[t]
	\includegraphics[width=\columnwidth]{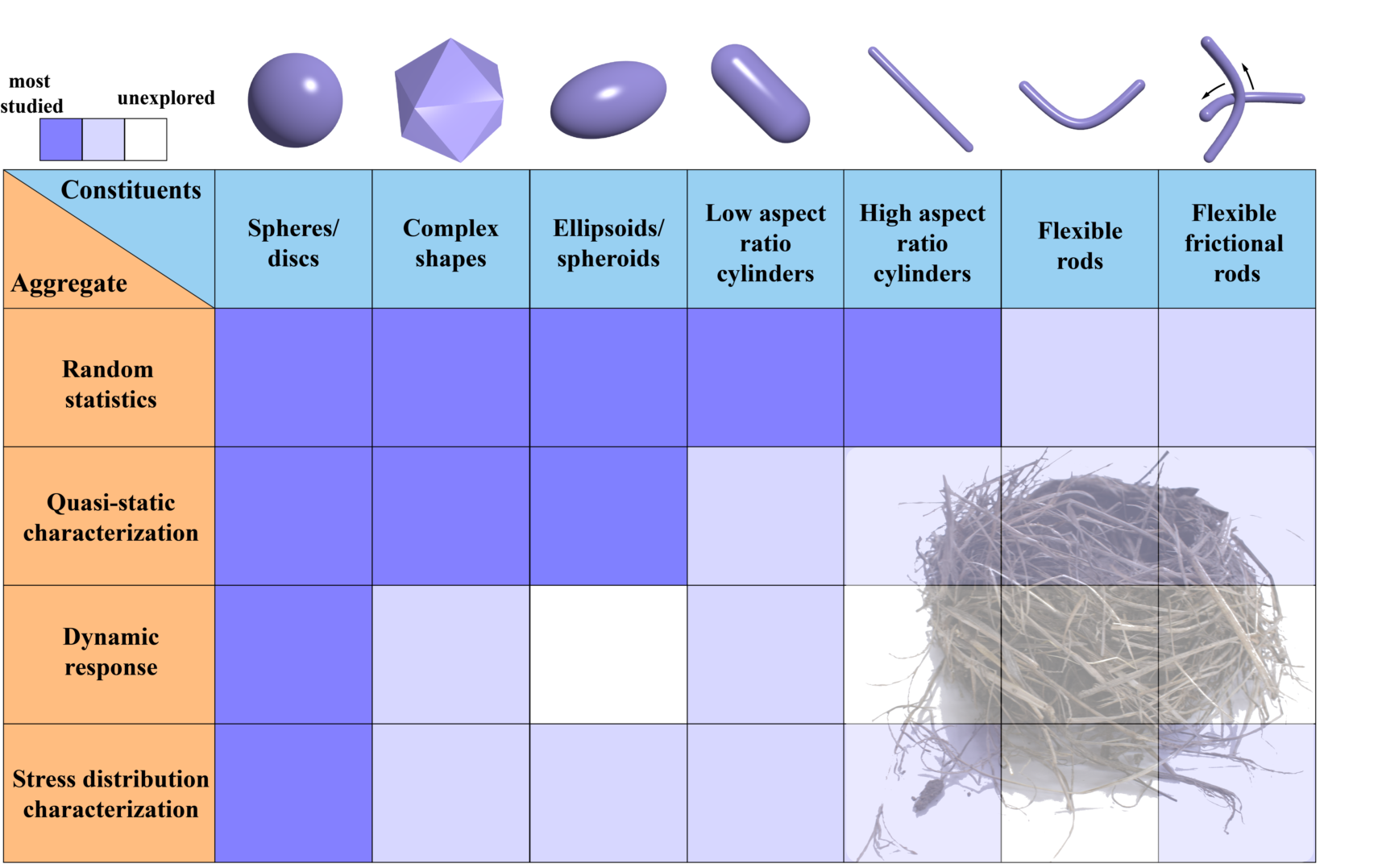}
	\caption{Crude depiction of the relative degree that aggregate mechanical behaviors of packed, athermal particles have been studied, for different types of constituent elements.  The shading corresponds to number of publications.  `Bird nest'-like systems occupy the unexplored region with flexible, frictional, high aspect ratio elements.}
	\label{fig:main}
\end{figure}

The most basic concept underlying nest stability, by which randomly packed grains come to behave collectively as a solid, has only a couple decades ago been given a scientific name: ‘jamming’\cite{ball1995lubrication}. This notion of a phase transition from fluid to disordered solid, in a system of athermal macroscopic elements, raises enough profound and fascinating questions to support a subfield of physics and an active line of research. The onset of jamming has been studied across diverse systems: foams, colloidal suspensions\cite{jam-phase-attractive}, and macroscopic elastic spheres\cite{jam-review,soft-jam}. 
A fundamental train of thought connects granular jamming to the glass transition\cite{liu}, and seeks to modify statistical mechanics to accommodate athermal analogs to characterize and predict the transition. Recent efforts have also been made to find generalities in jamming as it happens, i.e. to understand how jamming occurs, so as to usefully control its appearance and disappearance by design\cite{jamming_design}.  We will not look at that closely here.  \textit{Instead, we will focus on statistically robust jammed states as tunable, versatile materials.} Indeed, the evolutionary value of the bird nest appears to be in the mechanical properties of its jammed state, specifically those emerging from a subtle interplay between geometry, elasticity and friction between its slender, flexible elements.  This presents opportunities for the development of light weight, compliant, shock absorbing materials made of recyclable components. 
If bird nest design is to be taken as inspiration for synthesis of a granular material with tunable mechanical properties, the role of flexibility, friction and boundary effects at high aspect ratio should be carefully considered. As can be seen in Fig \ref{fig:main}, materials at the intersection of these features remain unexplored, prompting us to ask the following questions:  
\begin{itemize}
\item How do slender grains randomly pack?  What does the robust, emergent state look like and what statistical parameters define it? 
\item How does that state mechanically behave? What is common among static and dynamic responses of granular packings, and how do they depend on grain properties? What is the role of flexibility in slender grains?
\item How do boundary effects and packing protocols, critical when constitutive elements have sizes comparable to the whole system, modulate mechanical performance? Can we take advantage of boundaries, rather than seeing them as disturbances to bulk mechanics?
\end{itemize}

Answers could lead to minimal models and rational design guidelines to create lightweight materials with prescriptive, novel mechanical properties. Here, we aim to capture closest points of contact between these questions and the relevant scientific literature, and emphasize gaps where useful insight should be found. 

\section{\label{sec:level2}Nests as Meta-Material between granular packings and textiles}

\subsection{Statistics of jammed state}

\begin{figure*}[th!]
	\vspace{-10pt}
	\includegraphics[width=\textwidth]{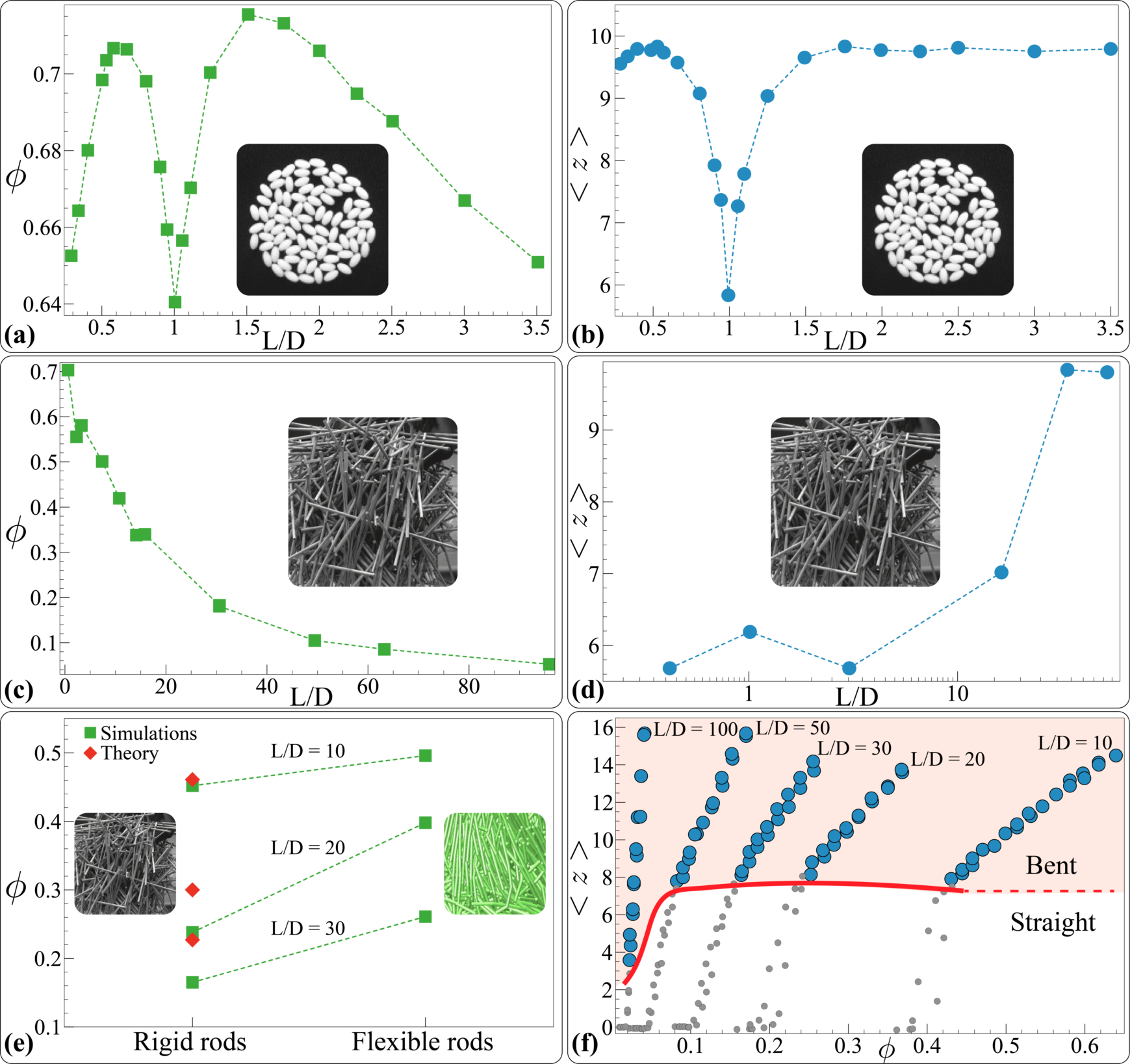}
	\caption{Donev et al.\cite{donev2004} explored the effect of varying the aspect ratio, using oblate and prolate spheroids (a and b) on packing fraction $\phi$ and coordination number $<z>$. The case of high aspect ratio stiff rods has also been well explored both numerically and experimentally\cite{fraden,philipse-contnum} (c and d). (e)DEM simulations demonstrate an increase in packing fraction for flexible rods when compared to  rigid rods\cite{langston2015discrete}. (f)  Coordination number increases beyond the rigid rod limit for flexible rods pointing towards an entanglement transition, as observed in DEM simulations\cite{rodney2005discrete}.}
	\label{fig:Donev}
	\vspace{-10pt}
\end{figure*}

Athermal grains lacking attractive interaction are capable of flowing around each other in response to external stress. If confined, however, they will readily `jam', i.e. assume properties of a continuous solid, mechanically identified by the onset of rigidity.
The jamming transition has been exploited for novel, practical application in recent years, as reviewed in Ref.\citenum{jamming_design}.  The ability to change a material from solid to fluid by mechanical, rather than thermal, actuation has been implemented in the grasping mechanism of a granular based robotic gripper \cite{brown2010universal} and in the propulsion of a soft robot \cite{steltz2010jamming}. In response to sudden compressive strain, granular systems have been reported to propagate dynamic jamming fronts which, by straddling the transition, dissipate significant energy, thus demonstrating their value as shock absorbers\cite{gomez2012uniform, gomez2012shocks}.
Potential value in the mechanics of the jammed state itself, and its tunability by variation of basic constituent characteristics has received less attention.  Particularly where that value and tunability derives from particle slenderness is the focus of our discussion.
	
In order to systematically characterize the structure and robustness of the jammed state, across the variety of grains, rods, and fibers considered, two statistical parameters are employed: (1) the volume fraction $\phi$, defined as the ratio between the grains' volume and total volume, measures how densely elements are packed; (2) the coordination number $<z>$, defined as the average number of contacts per element, measures mechanical connectivity of the aggregate. Values of these parameters depend on the system's confinement, preparation, and particle properties. Nonetheless, consistent trends across varying preparations are observed\cite{Scott_1969}, demonstrating the role of self-assembly in the disordered packing, without which the characterization of generic jammed states would be tenuous. We now look at these trends across preparations of (1) hard, spherical and spheroidal particles, (2) high aspect ratio, stiff rods and (3) extremely slender and/or soft fibers typically associated with textiles, to identify similarities and departures, and contextualize the nature of `bird-nest' materials.
	
Spheres have naturally received the most attention of the three cases.  
For the most dense regular (crystalline) packing, $\phi$ and $<z>$ take values of exactly 0.74 and 12, respectively\cite{Scott_1969}. Lack of orientational order, perhaps surprisingly, doesn't imply an arbitrary range of $\phi$ and $<z>$.  In fact, two distinct limits can be identified: random loose packed (RLP, $\phi\sim$ 0.55)\cite{RLP} represents the limit of minimal mechanical stability under gravity; and random close packed (RCP, $\phi$=0.64)\cite{Scott_1969} is the densest state for which spheres can pack without crystallizing. Where an aggregate of spheres finds itself between these limits depends on friction and on how gently/vigorously it is packed, but not sensitively on the details of the protocol\cite{greg}.

Particle shape can play a role in both the geometry of the packed state as well as in the kinetics of the packing process.  Slightly breaking spherical symmetry causes confined particles to rearrange by exerting torques on each other. Indeed, slightly oblate and prolate spheroids have been found to significantly deviate from spheres in packing fraction and coordination number.  Results from simulations (Fig. \ref{fig:Donev}a) show that as aspect ratio varies in either direction from unity, packing fraction first increases, reaching a maximum ($\phi \sim 0.72$) for moderately deformed grains, before trending downward\cite{donev2004}.  The coordination number (Fig. \ref{fig:Donev}b) is similarly minimal for a sphere (for which $<z> \sim 6$), but increases to a steady value of about 10 in either direction of aspect ratio.  The results show that for increasingly prolate particles, packings become less dense, while the average number of contacts per particle necessary to prevent further compaction remains constant.

Intuitively, this behavior should, to some degree, extend to stiff cylinders of increasing aspect ratio.  Experimental results for tinned buss wire, as shown in Fig.\ref{fig:Donev}c,d\cite{fraden}, are qualitatively consistent. After some discrepancy at aspect ratios $\lesssim$10, where edge effects are important, the number of contacts sufficient to maintain the jammed state saturates to $<z> \sim$ 10. The results agree with the geometry-based prediction of Ref.\citenum{philipse-contnum}, in which $\phi$ is dependent only on the aspect ratio.  
They have also been supported by numerical simulations\cite{williams2003random}, and appear to be generic (i.e. not strongly dependent on boundaries and packing protocol) as long as grain bending is negligible and pathological self-alignment is prevented \cite{villarruel2000compaction}. 

At some ticklish point in increasing aspect ratio and/or flexibility, one intuitively wants to call the system `fibrous' rather than `granular'.  While the transition point is far from clear, it would represent a bridge between two very different conceptual frameworks and language.  Viewed from the perspective of network mechanics, a fibrous system could be considered a disordered and `entangled' subset\cite{picu-review}.  In that context, `entangled' distinguishes their impermanent, frictional contacts from permanent chemical bonds. `Entangled' can also alternatively refer to the continuous overlapping of extended particles' rotational volumes\cite{wierenga1998low}, or the interpenetration of non-convex particles\cite{gravish}, which in either case inhibits their ability to rotate and rearrange with external stress.

%However in the case of flexible rods or fibers, entanglement can also occur due to the presence of bending mode deformations, which were originally negligible in the case of stiff rods.

%Toward this limit, we look at r
Raw textile materials, whose constituent elements have aspect ratios three or four orders of magnitude larger \cite{poquillon2005experimental} than the sticks of Fig. \ref{fig:Donev}c-d, clearly belong to the fibrous limit.  Reducing to this smaller, slenderer scale has two effects:  bending rigidity drops to the point where packing itself introduces bending\cite{asad2015characterization, hearle2008physical}, unlike experiments with rigid rods; and friction becomes increasingly important compared to gravity/inertia, making it difficult to reach RCP by successive vertical excitations.  The latter issue makes comparison of asymptotic packing statistics ambiguous, so that pure granular and textile literature are somewhat disconnected.  Nonetheless, a few studies, mostly computational, underscore potential bridges. 

%Figure \ref{fig:Donev}e,f illustrate similarities and differences between random packings of stiff sticks and flexible rods. 
An increase in packing fraction can be seen in Fig. \ref{fig:Donev}e, for Discrete Element Method (DEM) simulations of rods with increasing flexibility, loaded under gravity \cite{langston2015discrete}.  The sticks constrained from bending show values in agreement with theory\cite{parkhouse1995random, philipse-contnum} for stiff rods. Introducing flexibility, unsurprisingly, allows rods to pack to volume fractions beyond those seen for rigid rods, as rods are able to locally reorient during packing. 

Figure \ref{fig:Donev}f reports numerical simulation results of contact number and volume fraction for flexible rods of varying aspect ratio under isostatic compression\cite{rodney2005discrete}.  Below the red line, rods pack and rearrange without activating bending modes.  Pushed beyond a critical volume fraction, however, finite bending energy is measured in the fibers, indicating onset of what they call `entanglement'.  Notably, the packing fraction at the transition is quantitatively consistent with rigid rod RCP values in Fig. 3c, but the contact numbers disagree with those of Fig. 3d, both quantitatively and in the trend with increasing aspect ratio.  
%In particular, the coordination number $<z>$ is charted as a function of the fiber aspect ratio and the packing fraction $\phi$, which increases on compression. As can be seen, for low aspect ratio rods (AR $\sim 10-50$), the fiber systems quickly converge to packing fractions ($\phi \sim 0.1-0.5$) and coordination numbers ($<z> \sim 6-8$) consistent with observations for rigid rods (fig. \ref{fig:Donev}c, d). 
Upon further compression beyond the transition, an additionally important role of flexibility is seen, as
%Then, if additional compression takes place, the system's response changes character and the coordination 
the contact number starts growing linearly with the packing fraction, beyond the rigid rod limit. 
%This linear growth is consistent across all considered aspect ratios and in agreement with theory \cite{rodney2005discrete}, although for high values (AR $\sim 100$), this growth is observed to start at much lower values of $<z>$ and $\phi$ than the expected values for the rigid rod system. 
%This observation follows from the fact that higher aspect ratio rods are more flexible and thus start bending at $<z>$ and $\phi$ values lower than the equivalent jammed state of rigid rods. This change in behaviour of $<z>$ and the absence of the jammed state limit on $\phi$ and $<z>$ points towards the emergence of a qualitatively different transition in the system when compared to jamming of an equivalent rigid rod system. This we believe is due to the emergence of bending deformations and the increase in fibre-fibre contacts. Therefore, where jamming restrained the rigid rod system to a certain dense state, adding additional modes of deformation (eg. bending) allows the system to overcome this limit and pack to a denser entangled state.

%From grains to fibers, 'bird nests' are a largely unexplored material which lie at an unclear transitional regime; a seemingly jammed state with added flexibility and a potential to entangle. An exploration of random statistics characterization using aspect ratio and coordination number could provide broad insight into how grains go from rigid rods to becoming fibers and how the jamming transition evolves based on bending modes and consequent emergent entanglement in the system. 
Granular physics extends the conceptual framework of statistical mechanics to athermal systems by describing many particle ensembles as materials.  In simply looking at how they pack under generic confinement, we see that sticks and fibers can be described by the same average parameters used to describe jammed spheres.  In that space, a nest-like material occupies a region of predictably low density and asymptotically high contact number.  The process that brings them there is dependent mostly on particle geometry; flexibility introduces another degree of freedom in further deformation of the aggregate.

\subsection{Mechanical response}
%\rev{``Entangled networks'' describe the fibers/rods from the soft matter perspective -- how to describe aggregate response due to frictional (vs. crosslinked, permanent) contacts!!}

\begin{figure}[t]
	%\scalebox{-1}[1]{\includegraphics[width=\columnwidth]{V1a edits/imgs/stressstrainfigv1.png}}
	\includegraphics[width=\columnwidth]{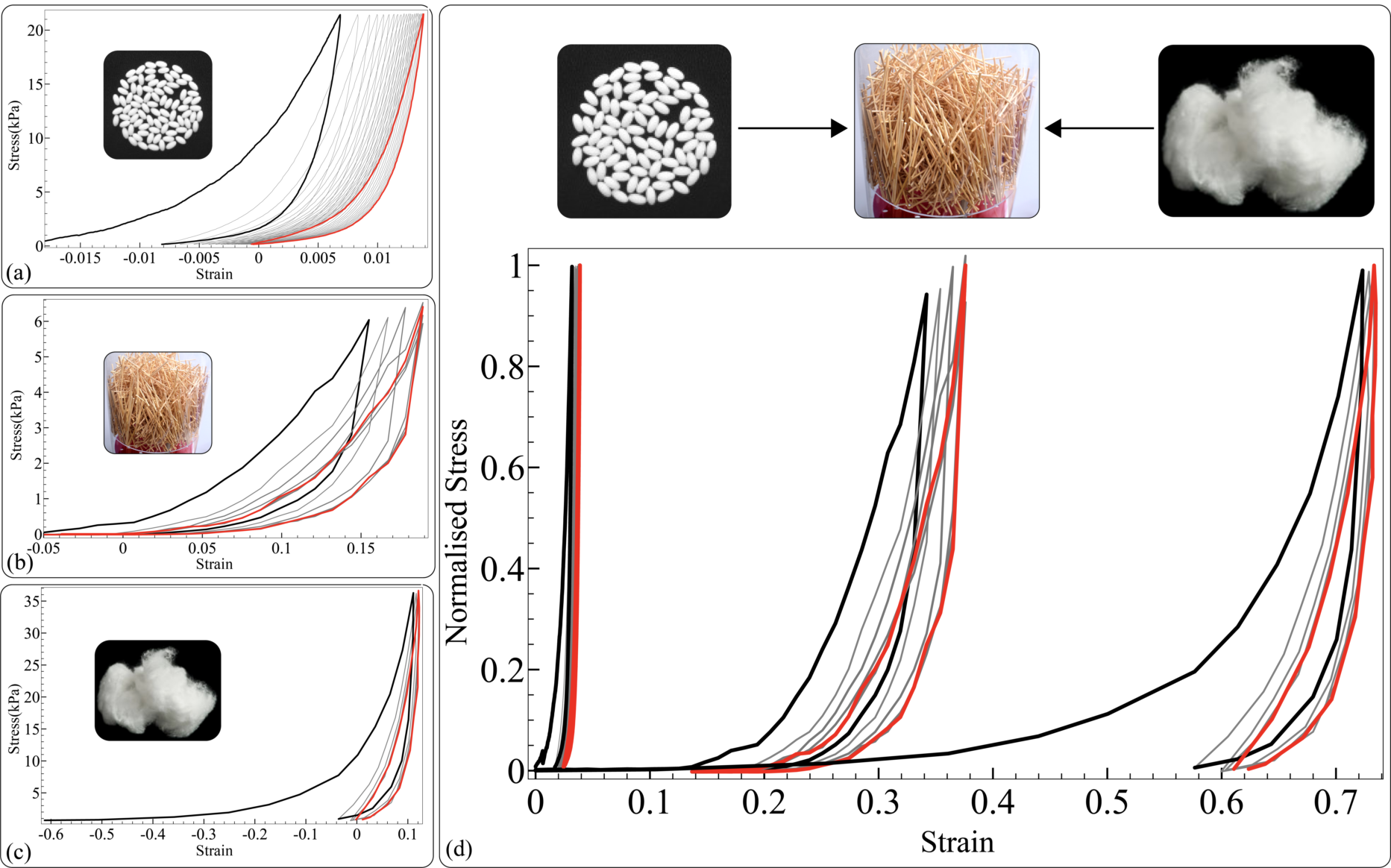}
	\caption{The same qualitative behavior can be seen in the stress strain data for (a) low aspect ratio spheroids \cite{lowaspect_stressstrain} (b) and high aspect ratio rods (c) and raw cotton fibers. \cite{poquillon2005experimental} (d) When plotted on the same normalized stress scales, the difference in plastic strain becomes evident. 
	}
	\label{fig:stressstrain}
\end{figure}

The previous section served to identify macrostates of the systems for which we anticipate reproduceable behavior.  We now explore simple mechanical characterization of those states, for packings of plastic `spheroids' of aspect ratio $\lesssim$2, wooden `rods' of aspect ratio $\sim$ 50, and natural raw cotton `fibers' with high aspect ratio $\sim$ 1000.

A recent study by Parafiniuk et al. looked at the stress response of packed spheroids to successive, cyclic, oedometric compression\cite{lowaspect_stressstrain}. The results, shown in Fig.~\ref{fig:stressstrain}a, reveal some noteworthy qualitative features.  % generic trends.  
The first cycle (bold black) loads and unloads non-linearly, carving out a large hysteretic loop and returns to a shifted zero-stress state.  Non-linearity is expected both because the coordination number (and force chain density) is likely increasing with deformation, and because the contact forces themselves are Hertzian\cite{johnson1987contact}. The subsequent cycles (gray) feature progressively decreasing plasticity, as the aggregate gradually compacts.  In this context, \textit{plasticity is a meta-material property}, in that it does not involve damage of primary material.  Instead, it results from rearrangement, attributed to decreasing orientational disorder \cite{lowaspect_stressstrain}, as the system moves from an initially somewhat loose packing (near RLP) formed during its preparation toward its random close packed (RCP) limit.  Here, between first and final cycle, the packing fraction increases from $\sim$0.54 %3 
to $\sim$0.56.%57 after a steady state cycle is reached. 
%Due to confinement effects, the asymptotic packing fraction at zero-stress state appearing in Fig.~\ref{fig:stressstrain}a are lower than the maximal packing fraction ($\phi=0.64$) reported by Ref.~\citenum{donev2004} (Fig.~\ref{fig:Donev}), but are roughly consistent.

When plastic deformation stops, finite hysteresis remains.  This persistent energy loss has been attributed to \emph{reversible} micro-slippage of frictional contacts\cite{2d_compression_spheres}.  The mechanism would be quasi-static, involving static friction.  As the aggregate loads and deforms, a given inter-particle contact experiences shear.  Upon overcoming static friction, the contact slides to a new equilibrium position.  The contact returns to its original positions as the load is relieved, but only after overcoming static friction in the opposite direction.  The return trip is less `springy' because the previous deformation is still temporarily stored in the network of frictional contacts.

For higher aspect ratio rods, flexibility plays a quantitative role in the mechanical response without changing the qualitative behavior from that of spheroids. Preliminary data for bamboo rods with aspect ratio 50 (Fig \ref{fig:stressstrain}b) shows initial plasticity in the first few cycles eventually giving way to a steady state cycle with hysteresis. The observed plasticity indicates that the initial state of the system was slightly looser ($\phi=0.071$) than the asymptotic value, and compacted under compression closer to that limit ($\phi=0.079$), which
%\rev{packing fraction starting from 0.0071 to 0.0079 for the steady state; the final packing fraction observed for the asymptotic zero-stress system
is consistent with values from Ref. \citenum{fraden} ($\phi=0.1$), but again lower due to confinement effects. Figure \ref{fig:stressstrain}d shows the data for spheroids\cite{lowaspect_stressstrain} and for our rods together, emphasizing the different scales of plasticity between the two systems.  Though the magnitude of applied stress differs significantly, the extent of final compaction appears not to depend on it: the respective asymptotic packing limits are approximately reached in either case.

Response of conventional fibers, with much higher aspect ratio $\sim$ 1000, to oedometric compression has been reported %in the context of textile and network mechanics literature\
by Ref. \citenum{poquillon2005experimental}. As shown in Fig \ref{fig:stressstrain}c, these display the same qualitative behaviors as those of both spheroid assemblies and slender rods: initial plasticity followed by a steady state cycle with hysteresis, a behavior first reported for textile materials by Van Wyk in the 1940s \cite{van194620}.  Again, increasing to more slender, more flexible fibers shifts the results quantitatively. 
%During preparation, the oscillations used to pack granular systems do not provide large enough
For small, fine fibers, gravitational and inertial forces are less sufficient to overcome frictional contacts, which leads to a much looser initial state, and therefore larger room for plasticity in compression.

%\NY{The following seems an independent point:}
Another investigation of fiber wads in uniaxial compression provides a micromechanical picture of the trend in Fig. \ref{fig:stressstrain}c, in two distinct steps.  First, fibers in an initially loose, random, 3D network bend and reorient away from the axis of compression, softly resisting with large plasticity.  By the end of the cycle, the system behaves as a 2D mat, largely aligned perpendicular to compression, displaying greater stiffness and low plasticity\cite{toll1998packing}.

Where friction at loaded, impermanent contacts appears to be the common feature of granular/fibrous systems which leads to their shared, distinctive plastic and hysteretic responses to successive quasi-static compression, slenderness appears to play a quantitative role.  
Increasing grain slenderness not only introduces a new (bending) mode of particle deformation, through which stresses are translated from contact to contact, it also introduces different time scales to the system.  Both should have consequences on the dynamic response of the aggregate.
%new deformation modes as one goes from grains to fibers additionally points towards the existence of new timescales appearing in the system.  This motivates the investigation of dynamic characterization of these systems.

%1.  Period of oscillation of a grain quickly changes in magnitude as it becomes slender -- xx for sphere, xx for rod, scaling with aspect ratio.
%2.  Force chains mediated by Hertzian contacts propagate stresses heterogeneously, but quickly, as each successive grain is in compression with the next.  By constrast, the unknown motif of stress propagation in fibers ... stresses propagate through bending modes, diffusive ...

For jammed grains to exhibit stiffness, stress must be propagated from grain to grain.  For spheres, stress is communicated through compressive deformations of successive grains in loaded contact.  This allows for an impulse to travel along a chain of contacts at speeds proportional to the square root of the stiffness of the Hertzian contacts\cite{Owens_2011}. For rods, a loaded contact exerts a torque on the portion of rod between itself and the nearest contacts which hold it in place.  A pulse, then, necessarily travels through many perpendicular paths mediated by the bending stiffness of the rod.  For very high aspect ratio, the distance between contacts becomes large relative to rod diameter $h$, and the bending stiffness decreases with $h^4$.  The combined factors must both lower the speed of propagation and increase the transverse diffusion of sudden stress.%, but to the best of our knowledge it has not previously been studied. 
The dissipation mechanism we described in the context of quasi-static compression is not inherently dependent on strain rate (as would be the case for a viscoelastic material), but rather originates from the sequence of local deformations which lock and release frictional contacts.  This implies that both storage and dissipation of energy from dynamic forcing would be non-trivially dependent on the frequency.
%Jammed spherical grains exhibit fast dynamics due to their Hertzian contact nature, and they have been shown to exhibit time scales of $O(10^{-6})$ s for mm scaled systems\cite{gomez2012uniform, gomez2012shocks}.  Dynamic responses of nest-like and fibrous systems have been less explored.  The stress relaxation response for mutton wool \cite{poquillon2005experimental} has found to be as slow as $O(10^2)$ s, which corresponds to the time scales of the bending mode of a geometrically equivalent flexible rod of AR $\approx$ 1000. 
At the same time, by manipulating the stiffness and geometry of slender rods, length- and time-scales could be injected in a controlled way,  
providing tunability of the dynamic response.  
%shock-absorbing and band-width filtering things.
%of the nest system away from both packed rigid grains and highly flexible fibers. 
%We speculate that in the case of flexible rods of moderate aspect ratio like the `bird nest' system, where the order of time scales of bending and Hertzian modes becomes less apart, an interesting and complex interplay between different modes may emerge. In this scenario, vibrations may be efficiently \rev{Yash: is efficiently the right word?} diffused via the long ranges enabled by the high aspect ratio elements.\rev{do we need a few more sentences about lengthscale?} This potential avenue of research is currently unexplored.

\subsection{Internal stresses and boundary effects}

\begin{figure*}[t]
	\includegraphics[width=\textwidth]{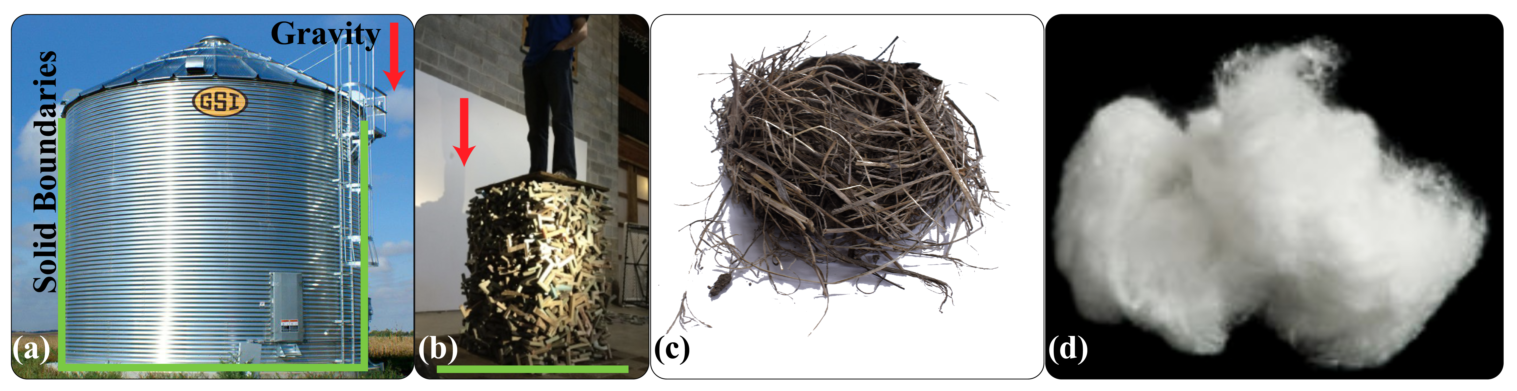}
	\caption{(a) Grains jammed by multiple boundaries under gravity in a grain silo \cite{silo}
	   (b)A structure jammed solely by the bottom wall and gravity made up of Z-form shapes \cite{murphy2016freestanding}
	   (c) A birds nest stable without any external forcing or boundaries
	   (d) A piece of cotton fiber stable without any external forcing or boundaries \cite{cotton}}
	\label{fig:be}
\end{figure*} 

%We are thinking of packed grains as `materials', which would imply they have intrinsic properties.  Because their jamming is induced by confinement, the mechanical response can't be completely separated from the role of boundaries.  
In the case of high aspect ratio elastic rods in a realistic nest, the notion of bulk properties becomes less clear and the role of boundaries becomes more central. In this context, how does the nest relate to grains and fibers?  For mostly round grains, the relationship between internal stresses and external boundaries was characterized in 1895, with Janssen's investigation of the fractional weight of grains supported by the walls of a silo (Fig \ref{fig:be}a)\cite{janssen}. Though the grains flow like a fluid as they are added to the container, the pressure at the bottom, rather than increasing linearly with the filling height of grains, saturates in the form of an exponential  $1 - e^{-\lambda z}$. The relationship between the vertical pressure $\sigma_{zz}$ and height $z$ from the top revealed a generic behavior dependent only on the parameter $\lambda$ or the decay length which depends on the pressure ratio $\kappa$, which arises from particle shape, wall geometry and friction.
	
The more recently developed force chain motif\cite{photoelastic_discs} complements this observation to complete an intuitive story.  As grains are added, contacts are formed until they meet the isostatic stability condition.  Increasing load on the disordered contacts focuses stress into heterogeneous 1D load structures between grains, which propagate through the material and end on the silo walls.  The end of the force chain exerts a normal force on the vertical wall, and through static friction relieves some of the total weight on the bottom.  Removal of the walls means breaking the force chains from their ends, losing the stability condition, and flowing -- in other words, unjamming.
	
This picture fails at some point for more exotic grain shape.  For extended particles with kinks, random packing causes interpenetration and entanglement, which complicate the stability condition by non-trivially coupling torques and normal contacts\cite{gravish, murphy2016freestanding}. For spheres, torques are mediated only by friction, but hooked particles must both rotate and translate in order to flow past each other.  Columns of hooked grains can be stable without walls, and can bear loads beyond their own weight, as shown in Fig \ref{fig:be}b. These columns and arches are reminiscent of sturdy platform nests, with contacts ultimately loaded by gravity.  The analogous stress-propagating motif for this more complicated case must still end at the lower external boundary to maintain force balance for gravity and for the existence of a stable non accelerating disordered interlocked structure.
	
Some cup nests, by contrast, can be turned upside-down, or even thrown against a wall without falling apart (Fig. \ref{fig:be}c).  This implies a jammed state that persists without any external confinement.  This situation is not easily compatible with the generic jamming mechanism described above, as there is no external force to load internal contacts nor a boundary for the analogous `force chain' to lean on.
One plausible explanation involves the additional role of flexibility in the construction process.  If sticks are forced to bend while packing, some of this bending stress could be stored in the system, held by frictional contacts. The resulting motif of stress propagation in the material would be highly complex, but it could, in principle, satisfy force balance with internal stresses, as a disordered version of a `stick bomb'\cite{sautel2017physics}.  This would place the bird nest in better company with unwoven textile materials such as fiberglass\cite{chen2016notch} or felt\cite{felt} -- disordered tangles of fiber whose mechanics derive from contacts loaded by internal bending stresses (Fig. \ref{fig:be}d).  When and how the force chain motif of the granular case, which requires boundaries for stability, is replaced by its analog for entangled fibers, in which a free floating sturdy structure is stabilized by bending modes remains an intriguing question for future research.

\section{Material design and role of simulations}

\begin{figure*}[t]
	%\scalebox{-1}[1]{\includegraphics[width=\columnwidth]{V1a edits/imgs/stressstrainfigv1.png}}
	\begin{center}
	\includegraphics[width=\textwidth]{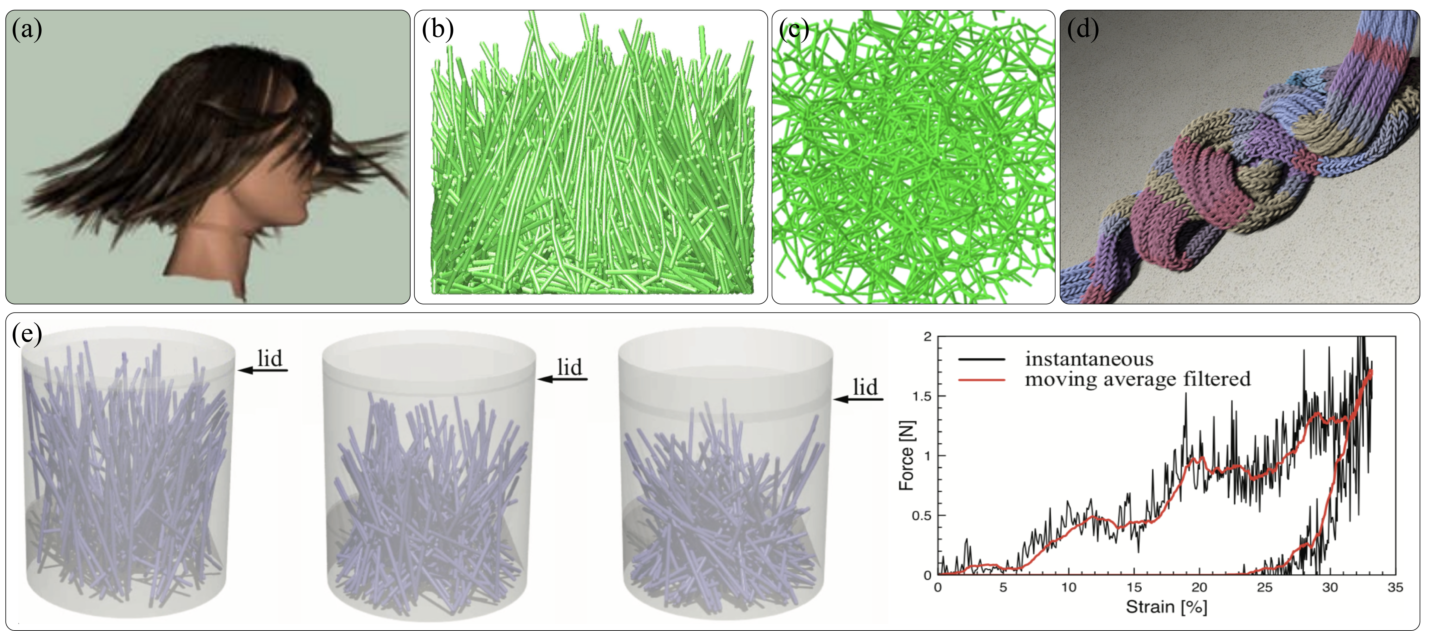}
	\end{center}
	\caption{Simulations of (a) natural hair \cite{bertails2006super}, (b) packing of flexible fibers \cite{langston2015discrete}, (c) fungal mycelium \cite{islam2017morphology}, (d) knitted cloth at yarn level \cite{kaldor2008simulating} and (e) a quasi-static compressive strain cycle for 250 rods and corresponding load-strain curve.}
	\label{fig:des}
\end{figure*}

As illustrated in the previous sections, `bird nest' systems may provide a route towards light weight, tunable and veratile materials. The key to harness these opportunities lies in establishing a systematic and efficient way to design packing systems (and confinement strategies) to meet desired target behaviours. Given the lack of theory guiding us, our poor physical understanding of the mechanisms at play, and the overall system complexity, computational modeling may prove  to be a powerful asset. Indeed, nest systems are characterized by a number of critical variables, from the geometric and material properties of their constituents to their boundary conditions. This vast space can hardly be explored experimentally. Hence, the need for predictive and efficient numerical models to complement and systematically integrate necessary experiments. Finite element methods have been employed for the simulation of disordered fiber systems \cite{islam2018stochastic} (Fig. \ref{fig:des}c). Although characterized by the highest level of fidelity, such methods are computationally expensive, rendering them impractical for the exploration of a vast design space. Alternative modeling approaches leverage the slender nature of fibers. These objects are then treated as one-dimensional systems, moving away from three-dimensional elasticity and significantly reducing the complexity of their mathematical representation. As a direct consequence, one-dimensional models are computationally inexpensive.  Several approximations have been proposed, from simple spring-mass systems able to capture stretching and bending modes at first order, to Discrete Rod Models \cite{kaufman2014adaptive} that accurately capture dynamics in three-dimensional space, accounting for various modes of deformation. The graphics community has been most active in this area, where the use of Cosserat model \cite{cosserat1909theorie} and its (far more popular) unstretchable and unshearable counterpart, the Kirchoff model \cite{kirchhoff1859uber}, have led to realistic simulations of elastic ribbons \cite{harmon2009asynchronous, bertails2006super}, woven cloth \cite{goldenthal2007efficient, kaldor2008simulating} (Fig. \ref{fig:des}d), entangled hair and fibers \cite{bertails2006super, durville2005numerical} (Fig. \ref{fig:des}a), wire mesh \cite{garg2014wire}, etc. These models also found application in physics, biology and engineering to characterize polymers and DNA \cite{yang1993finite, wolgemuth2000twirling}, flagella \cite{ko2017modeling}, tendrils \cite{gerbode2012cucumber}, cables in automotive design \cite{linn2017kinetic}, soft robot arms \cite{armanini2017elastica}, and dynamic musculoskeletal architectures \cite{gazzola2018forward, pagan2018simulation, charles2019topology, aydin2019neuromuscular, xz2019}.

 These representations are versatile and accurate, can be easily interfaced with dynamic environmental loads (contact, friction, hydrodynamics), and significantly reduce complexity and computational costs, in particular relative to standard approaches based on finite element methods. Figure \ref{fig:des}e illustrates that this approach can be adopted to the study of nest systems. There, we considered thin, virtual wooden sticks characterized by circular cross sections and with bending stiffness comparable to the setup of Fig. \ref{fig:stressstrain}b. Friction among sticks was assumed to be isotropic and estimated through friction tests. Contacts were also detected and accounted for through a repulsive force. Finally, for simplicity, the sticks-container interaction was modeled as the stick-stick interaction. We then simulated a cycle of compression and recorded the `nest' response on the lid of the container. As can be noticed in Fig. \ref{fig:des}e, the generated load strain curve qualitatively captures the highly hysteric behaviour of Fig. \ref{fig:stressstrain}b. We note that because of the limited number of rods employed (250), the load measurements are noisier than the experiments, nevertheless the trend is clearly captured. 

These techniques thus can be employed to ``preview'' in-silico (forward design) dynamic behaviour of a given aggregate. At the same time, given their moderate computational costs and numerical robustness, they are also suitable to perform inverse design tasks, in which key properties of the aggregate are identified in order to achieve an overall target system's behaviour. In this context, the use of evolutionary strategies is particularly promising. These class of algorithms generate populations of candidate solutions (i.e. different nest systems). These are simulated independently from each other (allowing us to distribute them across supercomputing facilities) and their dynamic response is evaluated according to a desired metric. The best solutions are then recombined, to generate a new, more performant pool of aggregates, until no significant improvement is observed. We have employed these procedures in combination to Cosserat models, to improve the locomotory performance of soft-robots\cite{aydin2019neuromuscular, xz2019, charles2019topology}, demonstrating the practical viability of this approach.

\section{Outlook}

\begin{figure*}[t]
	%\scalebox{-1}[1]{\includegraphics[width=\columnwidth]{V1a edits/imgs/stressstrainfigv1.png}}
	\begin{center}
	\includegraphics[width=\textwidth]{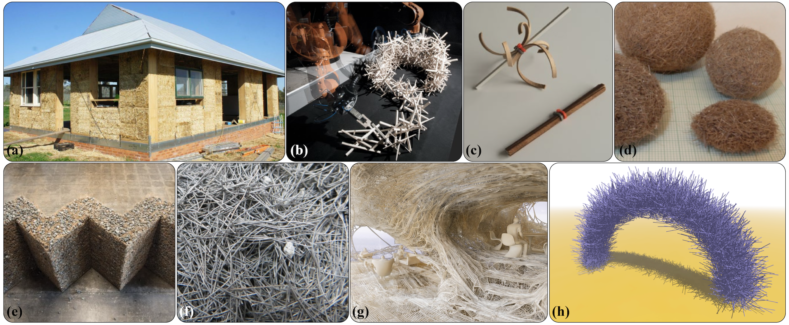}
	\end{center}
	\caption{(a) Straw-bale based construction\cite{straw}, (b) Adaptive structure built using interlocking particles, \cite{dierichs2017granular}
	(c) Particles changing shape based on a hygroscopic response\cite{dierichs2017granular}, (d) Aegagropilae, a entangled fiber network formed from dead sea grass in the ocean \cite{verhille2017structure} (e) Stable granular structures built using rope as internal confinement\cite{Kholer},
	(f) SEM image of fiber reinforced concrete\cite{frc}, (g) Fiber based architectural design of a modern house\cite{fhouse},
	(h) Adaptive, flexible and tunable arch made of granular flexible rods.}
	\label{fig:apps}
\end{figure*}

%The 'bird nest' may have similarities to many of these systems but is also fundamentally different. 

%A system which presents itself at the intersection of granular media physics and random fiber network physics we believe has broad potential in the development of structures that can jam-unjam partially, absorb impact in unique ways, possess the strength of entangled fibers while adapting to the surroundings based on constituent properties and packing protocol(Fig \ref{fig:apps}h). 
%We believe these structures could have universal applications; acting as a jammed construction material at one instance, while flexibly moving to perform a delicate surgery at another, as well as existing as a stable structure even without external confinement in places like the outer space.}
This perspective paper aims at contextualizing nest-like aggregates as materials, characterized by rich dynamic behaviors at an unknown transition between classic hard grains and textile fibers.  As such, they may provide an avenue to bridge the range of applications typically associated
with these apparently disparate systems.

Much work has focused on the drivers, implications, and characteristics of the jamming transition in grains.  
%The transition of the unjammed to jammed state for spheroidal particles and complex shapes with low aspect ratios has been applied to a variety of applications, such as
Manipulation of the transition, back and forth between fluid-like and solid-like properties via external confinement has presented a basis of actuation in soft robotics \cite{steltz2009jsel, brown2010universal, loeve2010vacuum} and deformable aerostructures \cite{aerostructures}.  Architects and artists have recently demonstrated the elegance and practical versatility that comes with embracing disorder and self-assembly instead of prescriptive control to build reconfigurable structures of emergent stability.  This has been demonstrated both through choice of particle geometry\cite{dierichs2016towards,dierichs2017granular,Keller2016,murphy2017aleatory} and composite granular/fibrous building material\cite{aejmelaeus2017granular,Kholer}.
Where these and other applications make use of when and how grains jam, the richness in their mechanical response to deformation has not been harnessed to similar effect.  By controlling slenderness and flexibility in even simple shapes, this richness can be explored.

On the other hand, understanding mechanics of entangled random fiber networks have been at the core of development in textiles\cite{choong2013compressibility,poquillon2005experimental}, ballistic impact mitigation\cite{roylance1977ballistics, cunniff2002high}, fiber based architecture (Fig. \ref{fig:apps}g)\cite{fhouse} and construction \cite{fcons}.  Use of compressed, randomly packed hay bales as building material (commonly termed as straw-bale construction, Fig. \ref{fig:apps}a)\cite{straw} has been revived as an environment friendly option for construction, providing structure and insulation. %This is similar to the use of a granular material to fill sandwich beams83 where it acts as a light weight filler with damping benefits. 
Naturally occurring examples will continue to provide insight toward intelligent application.  Under water, bouyancy and random flows can cause fibrous structures to passively self-assemble.  Aegagropilae (Fig. \ref{fig:apps}d) -- balls of dead, sea grass which are gradually entangled and packed by random ocean currents -- appear mechanically very similar to continuous, cohesive, elastic spheres.  The scaling of their effective modulus with mean density has been effectively modeled based on filament bending stiffness and contact number\cite{verhille2017structure}.

Real bird nests have inspired scientific study for hundreds of years, but the underlying logic from a practical, physical perspective is coming closer to focus with research into both nest structure and building behavior.  While several forces have been shown to drive bird nest design \cite{structural_morph,design-funct,latitude,humidity} across the diversity of nests\cite{hansell-book,bird_engineering}, the need for structural integrity under mechanical loads and disturbances, over the lifetime of the nest, is seen as a dominant factor\cite{struct_supp_nests}.  
Some of the intense complexity of weaver nest morphology has been found to emerge, in part, from processes involving simple construction rules and self-organizing mechanisms such as stigmergy\cite{const-rules}.  Observations of the nesting behavior in zebra finches show selection criteria based on filament geometry and flexibility\cite{muth2014zebra,bailey2014physical}.  Further cross-disciplinary study of bird nesting behavior, with controlled material inputs, output mechanical characterization, and complementary experiment and simulation of artificial analogs, could reveal generalizable construction algorithms of significant biological and technological impact.

Between sand and cloth, at a blurry interface between granular and textile mechanics, exists a class of material familiar to the biological world but relatively unexplored with scientific rigor.
%may provide a versatile mechanical framework with significant potential in the development of structures that can jam-unjam partially, absorb impact in unique ways, possess the strength of entangled fibers while adapting to the surroundings (Fig 7h).
%The understanding and characterization of collective properties of nest materials made of jammed filaments may therefore 
Understanding this region of parameter space could generate new points of theoretical traction into granular physics, illuminate functional mechanisms in animal engineering that extend beyond birds and biological curiosity, and inform design of lightweight materials with prescriptive mechanical properties that cut through many areas of high current importance: civil engineering and architectures (reliable, inexpensive, reusable and self- repairing construction materials), transportation (lightweight composites, shock absorbers), and advanced manufacturing.

\bibliography{gran}
	
\end{document}